\documentclass[11pt]{JHEP3}
\usepackage{graphicx}
\usepackage{epsfig}
\usepackage{amssymb}
\usepackage{amsmath}
\input epsf.tex

\newcommand{\gsim}{\mbox{\raisebox{-1.ex}{$\stackrel
      {\textstyle>}{\textstyle\sim}$}}}

\newcommand{\half}{\ensuremath{\frac{1}{2}}}

\newcommand{\be}{\begin{equation}}
\newcommand{\ee}{\end{equation}}
\newcommand{\bea}{\begin{eqnarray}}
\newcommand{\eea}{\end{eqnarray}}

\newcommand{\ns}{\normalsize}

\title{Particle transfer in braneworld collisions}

\author{
  Paul.M.Saffin$^{1}$\footnote{email: paul.saffin@nottingham.ac.uk} and
  Anders Tranberg$^2$\footnote{email: a.tranberg@damtp.cam.ac.uk},\\
  $^1${\it\ns School of Physics and Astronomy, University of Nottingham}\\
  {\it\ns University Park, Nottingham NG7 2RD, United Kingdom.}\\  
  $^2${\it\ns DAMTP, University of Cambridge} \\
  {\it \ns  Wilberforce Road, Cambridge, CB3 0WA, United Kingdom.}
}

\keywords{Kinks, fermions, branes}   

\preprint{DAMTP-2007-47}

\date{}

\maketitle

\abstract{We study the behaviour of fermions localized on moving kinks
as these collide with either antikinks or spacetime boundaries. 
We numerically
solve for the evolution of the scalar kinks and the bound
(i.e. localized) fermion modes, and calculate the number of fermions
transfered to the antikink and boundary in terms of Bogoliubov coefficients. 
Interpreting the boundaries as the brane on which we live, this models 
the ability of fermions on branes incoming from the bulk to ``stick'' on 
the world brane, even when the incoming branes bounce back into the bulk. }

\begin{document}



\section{Introduction\label{sec:introduction}}

Fermions coupled to inhomogeneous background bosonic fields can become
localised, in the sense that the fermion spectrum contains localised
bound states
\cite{Jackiw:1975fn,Jackiw:1981ee,Niemi:1984vz}.
In particular in the presence of scalar kinks, fermions
prefer ``living on'' the kinks rather than populate higher energy
(delocalised) radiation modes. This phenomenon is ubiquitous in
condensed matter systems with impurities and where external magnetic
(gauge) fields provide the inhomogeneities. The case of
scalar fields acquires particular interest, if one allows kinks to
represent higher dimensional domain walls or branes in string
theory \cite{DeWolfe:1999cp}. The localisation of fermions on inhomogeneities then becomes
reminiscent of brane world scenarios, where the Standard Model fields
are expected to ``live'' on one particular three dimensional brane. 

In the presence of extra dimensions, additional branes may exist (in the ``bulk''), 
each with their localised fermion states. As such branes collide, one may envisage 
fermions being transfered from one brane to another. Such a scenario was studied in
\cite{Gibbons:2006ge}.

Another picture of braneworlds emerges from the work of Horava and Witten \cite{Horava:1996ma},
where it was realized that spacetime boundaries can play an important role. In that
particular case it was discovered that each of the two boundaries supported an E8
gauge theory. This braneworld model then places our Universe at the boundary of some
larger spacetime. The question of what happens as bulk branes collide with our
boundary-Universe is what this paper addresses, as well as extending the calculation of 
Gibbons {\it et al} \cite{Gibbons:2006ge}.

In order to model a 3+1 Universe as a boundary of a 4+1 spacetime we could start with
an action of Dirac fermions, and scalar fields that support
domain walls to model the branes. However,
if we employ a planar symmetry along the walls this model reduces to a 1+1 model with a boundary.
And for our particular initial conditions (domain walls with just a bound zero mode) the
equations for the Dirac fermion reduce to the equations for a Majorana fermion. So, while we in
practise simulate a Majorana fermion in 1+1, this can be lifted to a 
4+1 spacetime with Dirac fermions,
a scalar field for the branes, and a boundary.

We study the fermion transfer numerically, treating the scalar as classical and the 
fermions in terms of a set of quantum modes.


\section{Scalar-fermion model in 1+1 dimensions\label{sec:model}}

We consider a model of a real scalar $\phi$ and a single fermion
species $\psi$ in 1+1 dimensional space-time.
In our simulations we shall be considering situations both with and without a boundary,
in either case the bulk action we use is
\bea
\label{eq:action}
S_{\rm bulk}&=&-\int dt\,dz\left[ \frac{1}{2}\partial_{\mu}\phi\partial^{\mu}\phi
                    -i\bar{\psi}\gamma^{\mu}\partial_{\mu}\psi
                    +\frac{\lambda}{4}\left(\phi^{2}-1\right)^{2}
                    -ig\phi\bar{\psi}\psi\right],
\eea
with the bosonic boundary action being
\bea
\label{eq:actionBndry}
S_{\rm boundary}&=&\mp\int dt   \left[ \sqrt{\frac{\lambda}{2}}\left(\frac{1}{3}\phi^3-\phi\right)
                               \right]_{z=0},
\eea
Our choice of conventions is
\bea
\eta^{\mu\nu}=diag(-1,1),\quad
\{\gamma_\mu,\gamma_\nu\}=2\eta_{\mu\nu},\quad
\bar\psi=\psi^\dagger\gamma_0.
\eea
At this point, the couplings $\lambda$ and $g$ are free
and we have chosen the boundary to lie at $z=0$.
The boundary action for the scalar may
seem unnatural at first but, as we shall see, such a term means that there is no force between the
boundary and a (static) kink, (or antikink if the + sign is chosen).
Another reason is that this bosonic boundary term is required for the action to be supersymmetric,
along with taking the spinor to be Majorana with couplings related by $g^2=2\lambda$. We shall not be
requiring supersymmetry.
The corresponding equations of motion in the bulk are
\bea
\left(\gamma^{\mu}\partial_{\mu}+g\phi(z,t)\right)\psi(z,t)&=&0,\\
\left[\partial_{\mu}\partial^{\mu}-\lambda\left(\phi^{2}(z,t)-1\right)\right]\phi(z,t)
          &=&-ig\bar{\psi}\psi(z,t),
\label{eq:phieom}
\eea
By choosing a real representation of $\gamma_{\mu}$,
\bea
\gamma_{0}=
\left(
\begin{array}{cc}
0&1\\
-1&0
\end{array}
\right)
,~~~
\gamma_{z}=
\left(
\begin{array}{cc}
0&1\\
1&0
\end{array}
\right).
\eea
the equation of motion for the complex two-component fermion splits up
into two uncoupled real (Majorana) copies,
\bea
\psi=\psi^{M}_{1}+i\psi^{M}_{2}.
\eea
The equations of motion for $\psi^{M}_{1}$ and $\psi^{M}_{2}$ are identical, 
and we will from now on think in terms of a single, two-component Majorana fermion 
$\psi_1^M$, ignore the superscript, and write it as
\bea
\psi^M_1(z,t) = \left(\begin{array}{c}\psi_{1}(z,t)+\psi_{2}(z,t)\\ \psi_{1}(z,t)-\psi_{2}(z,t)\end{array}\right).
\eea
The equations of motion then read
\footnote{These equations are equivalent to those 
of the 4+1 system in \cite{Gibbons:2006ge} if we take
$\psi_1=i\psi_+$ and $\psi_2=\psi_-$. This explains why the
authors see an additional phase of
$\pi/2$ between $\psi_+$ and $\psi_-$.}

\bea
\label{eq:eom}
\dot{\psi}_{1}(z,t)&=&-\partial_{z}\psi_{2}(z,t)+g\phi\psi_{2}(z,t),\\
\label{eq:eom2}\dot{\psi}_{2}(z,t)&=&-\partial_{z}\psi_{1}(z,t)-g\phi\psi_{1}(z,t),\\
\ddot{\phi}(z,t)&=&\partial_{z}^{2}\phi(z,t)-\lambda\left(\phi^{2}(z,t)-1\right)\phi(z,t)
\eea
We discretise these in a straightforward way and solve them
numerically using a standard algorithm, 4th. order accurate in time,
2nd order in space. The boundary condition was satisfied using a
Newton-Raphson iteration \cite{Antunes:2003kh}.

The boundary conditions coming from the action are
\bea
\label{eq:boundarycondBoth}
\partial_{z}\phi|_0&=&\mp \sqrt{\frac{\lambda}{2}}\left(\phi^{2}-1\right)|_0,
\eea
and for the fermions we will be using 
\be
\psi_1|_0=\mp\psi_1|_0,\;\psi_2|_0=\pm\psi_2|_0,
\ee
which can be derived from a boundary action of $\pm\frac{i}{2}\int\;dt[\bar\psi\psi]_{z=0}$.
We shall call these the $\mp$ Boundary Conditions, $\mp BC$.
We see therefore that at the boundary
either $\psi_1$ or $\psi_2$ vanishes, depending on the
choice of sign of boundary conditions, with the
$-BC$ giving $\psi_1|_0=0$ and the $+BC$ giving $\psi_2|_0=0$. As we shall see later, these boundary
conditions allow for a normalizable fermion condensate on the boundary to co-exist with a kink
or antikink.

The scalar field equation can be
thought of as concerning a classical field, but as fermions are 
quantum, the right
hand side source term should be replaced by
$-ig\langle\bar{\psi}\psi\rangle$. The resulting equations amount to a quantum
fermion in a classical scalar background.

For the purpose of this paper, we will make the approximation of neglecting the fermion
back-reaction on the scalar. 
The upshot of this approximation is that whereas the solution of
the mode equations (see below) are independent of the initial state
(i.e. particle content) of those modes, the back-reaction term
$\langle\bar{\psi}\psi\rangle$ is a quantum average over some density
matrix. In the present
 context, we are more interested in the behaviour of the fermion modes
 as the scalar background changes than the exact details of the kink
 evolution. It is however clear that with many fermions present, in
 particular when including all the non-localised modes, the
 back-reaction may be sizeable, and may even drive the system to a
 high-temperature thermal state. Then the setup of a solitary kink or
 kink-antikink pair may no longer be realistic. 

In our simulations we monitor the conservation of energy and we give here the expression
of the energy in the scalar field,
\bea
H&=&\int dx\left[\half(\dot\phi)^2+\half(\phi')^2+\half\left(\frac{dW}{d\phi}\right)^2\right]\pm W(z=0).
\eea
In particular we see that the boundaries have an energy associated with them. In the case where the
scalar field is in the vacuum at the boundary we see that the boundary has an associated energy
of 
\bea
\label{eq:bndryEnergy}
E_b(\phi=1)=\mp\frac{2}{3}\sqrt{\lambda/2},~~~~~E_b(\phi=-1)=\pm\frac{2}{3}\sqrt{\lambda/2}.
\eea
This boundary energy will be important in understanding the dynamics of kink-boundary collisions.


\subsection{Kinks and boundaries\label{sec:kinksboundaries}}

The static scalar equation of motion has a kink and an antikink solution
$\phi_{K}$ and $\phi_{A}$,
\bea
\phi_K(z)= \tanh\left(\frac{z-z_{0}}{D}\right),~~
\phi_A(z)=-\tanh\left(\frac{z-z_{0}}{D}\right),~~D=\sqrt{\frac{2}{\lambda}},
\eea
where $z_{0}$ is the center of the (anti-)kink. 
We will use $\lambda=2$, $D=1$ throughout.\footnote{This amounts to rescaling
the original action (\ref{eq:action}).}
Note that these static solutions obey
\bea
\label{eq:boundarycond}
\partial_{z}\phi=\mp \sqrt{\frac{\lambda}{2}}\left(\phi^{2}-1\right),
\eea
so that there is no force between a kink and a $-BC$. Similarly there
is no force between an
antikink and the $+BC$. In the case where the coupling constants take the supersymmetric values,
$g^2=2\lambda$, the (anti)kink and the ($+BC$)$-BC$ break the same half of the supersymmetry.

When colliding kink with antikink we shall employ periodic
boundary conditions. When colliding single kinks onto a boundary,
these are placed at $z=0$, with the kink coming in from the left $(z<0)$. 
Note that there is no loss of generality by choosing to send in kinks
but not antikinks, while using both boundary conditions. The system of
kink and $-BC$ is equivalent to antikink and $+BC$, while kink and $+BC$ is equivalent
to antikink and $-BC$; this covers all the possibilities.


\section{Bound states\label{sec:fermbound}}


\subsection{Static kinks}
An interesting facet of topological defects such as kinks is that they often allow for bound
states in the particle spectrum. In the case at hand there are bound states for both the scalar
and fermi fields. Indeed, the existence of the scalar bound state was the motivation behind the
work of Rubakov and Shaposhnikov \cite{Rubakov:1983bb}, asking whether we live on a
domain wall. To study the scalar spectrum we consider perturbing the scalar equation of
motion in (\ref{eq:phieom}) around a kink
\bea
\phi(z,t)&=&\phi_{\rm kink}(z)+\delta\phi(z,t),
\eea
then by writing
\bea
\delta\phi&=&\exp(i\omega t)F(z),
\eea
we solve the resulting eigenvalue equation to find the frequencies of the bound states
\cite{Dashen:1974cj,Rubakov:1983bb}. Taking the change of variables \cite{Gibbons:2006ge}
\bea
Z=\tanh(z/D),
\eea
we arrive at an associated Legendre equation with $l=2$,
\bea
(1-Z^2)\frac{d^2 F}{dZ^2}-2Z\frac{dF}{dZ}+2(2+1)F-\frac{(4-\omega^2D^2)}{1-Z^2}F&=&0,
\eea
for which there are three solutions, $P^2_{\;2}(Z)$, $P^1_{\;2}(Z)$, $P^0_{\;2}(Z)$,
corresponding to $m^2=4-\omega^2 D^2=2,\;1,\;0$. These give frequencies $\omega=0,\;\sqrt{3}/D,\;2/D$.
The first of these modes, the zero mode, corresponds to the translation of the kink, the second is a true
bound state while the third is not normalizable and so is not considered to be in the physical spectrum. We also
note that this last solution is right on the boundary of the continuum states, which have mass $\sqrt{2\lambda}=2/D$.

As well as these scalar bound states there are fermion modes localized to the kink which we now describe.
First there is the zero-mode \cite{Jackiw:1975fn}
\bea
\psi^{K}_{1}(z,0)=\sqrt{\frac{\Gamma[gD+1/2]}{2D\sqrt{\pi}\Gamma[gD]}}\frac{1}{\cosh[(z-z_{0})/D]^{gD}},~~~\psi^{K}_{2}(z,0)=0,
\eea
where we have normalized too unity according to the inner product
\bea
\label{eq:fermiNorm}
(\psi,\chi)&=&\int dx\psi^\dagger\chi.
\eea
Now recall that the scalar field of the kink can happily co-exist with the -BC, and 
that these boundary conditions required the vanishing of $\psi_2(z=0)$. So we see that the -BC
also have no effect on the fermion fields of the static kink.

In addition to the zero mode we also find that there is an excited fermion mode given by
\bea
\label{eq:Emodes1}
\psi^{KE}_{1}(z,t)&=&-\mathcal{N}\omega D
\frac{\sinh\left(\frac{z-z_{0}}{D}\right)}{\left[\cosh\left(\frac{z-z_{0}}{D}\right)\right]^{gD}}\cos(\omega
  t-\varphi),\\
\psi^{KE}_{2}(z,t)&=&\mathcal{N}\frac{1}{\left[\cosh\left(\frac{z-z_{0}}{D}\right)\right]^{gD-1}}\sin(\omega
  t-\varphi),
\label{eq:Emodes2}
\eea
with
\bea
\omega^{2}=\frac{2gD-1}{D^{2}},~~~
\mathcal{N}^{2}=\frac{\left(gD-1\right)\Gamma[gD+1/2]}{\left(2gD-1\right)D\sqrt{\pi}\Gamma[gD]}.
\eea
Recall that this model, with a Majorana fermion, is supersymmetric when $g=\sqrt{2\lambda}=2/D$.
In this case we see that
the frequency of the excited fermion mode matches that of the scalar bound state, $\omega=\sqrt{3}/D$,
as is expected for supersymmetry. We also expect, given that there are no more scalar bound states, that there
will be no more fermion excited states. We note that this does not hold true as we change couplings
away from the supersymmetric case.\footnote{We would like to thank Kei-ichi Maeda 
for informing us of his calculation showing that there is a tower of bound
states with frequency $\omega^2=n\frac{2gD-n}{D^{2}}$, for integer $n$ subject to $0\leq n <gD$.}


\subsection{Moving kinks}
As we are interested in kinks colliding with each other and against
boundaries we need to know what kinks and bound states look like when
they are moving, that is, we need to boost the kink.

The scalar field of a kink (antikink)
moving at speed $v$ simply involves a Lorenz contraction by $\gamma=1/\sqrt{1-v^2}$,
\bea
\phi^{K/A}_{v}(z,0)&=&\pm\tanh\left[\gamma\left( \frac{z-z_{0}}{D}\right)\right].
\eea

An isolated, moving kink or antikink has an energy of
\bea
\label{eq:kinkEnergy}
E_{\rm antikink}=E_{\rm kink}&=&\frac{4}{3}\sqrt{\lambda/2}\gamma.
\eea
We note that the energy of a static kink, $v=0$ coincides with the
difference in energy between a $-$ and a $+$ boundary, $\Delta E_{\rm
  b}=\frac{4}{3}\sqrt{\lambda/2}$, from Eq. (\ref{eq:bndryEnergy}).

The fermion modes also transform in the usual way, and in terms of the
$1,2$ components we have
\bea
\label{eq:boostedmodes}
\psi_{1}(z,t)\rightarrow\psi_{1,v}'(z,t)&=&\sqrt{\frac{\gamma+1}{2}}\left(
\psi_{1}\left[\gamma (z-vt)\right]+\frac{v\gamma}{\gamma+1}\psi_{2}\left[\gamma (z-vt)\right]\right),\\
\psi_{2}(z,t)\rightarrow\psi_{2,v}'(z,t)&=&\sqrt{\frac{\gamma+1}{2}}\left(
\psi_{2}\left[\gamma (z-vt)\right]+\frac{v\gamma}{\gamma+1}\psi_{1}\left[\gamma (z-vt)\right]\right).
\eea


\subsection{Boundaries}\label{sec:onboundary}

Just as kinks support bound states, so does a boundary. To describe these we consider the
scalar field to be in one of its vacuum states at the boundary, $\phi(z=0)=\pm 1$,
then we see from the
spinor equations of (\ref{eq:eom}) that the boundary
carries a single localized, normalisable fermion
\bea
\partial_{z}\psi_{1}(z)=-g\phi(z)\psi_{1}(z),~~~
\partial_{z}\psi_{2}(z)=g\phi(z)\psi_{2}(z).
\eea
For $\phi(z=0)=+1$ the normalisable solution is given by
\bea
\psi_{1}^{B}(z)=0,~~~\psi_{2}^{B}(z)=\sqrt{g}\exp\left(gz\right).
\eea
For $\phi(z=0)=-1$ we instead have
\bea
\psi_{1}^{B}(z)=\sqrt{g}\exp\left(gz\right),~~~\psi_{2}^{B}(z)=0.
\eea
Note that the condition of normalisability imposes that one fermion
component is zero in each case.
For example, if we have a single kink in the bulk (which allows
a condensate of $\psi_1$) then at the boundary
we will have $\phi\simeq +1$ which allows a boundary condensate of $\psi_2$,
consistent with the -BC.


\section{Kink dynamics\label{sec:kinkdynamics}}

\FIGURE{
\epsfig{file=./pictures/KK_NR_csample.eps,width=6cm,clip}
\label{fig:confKAK}
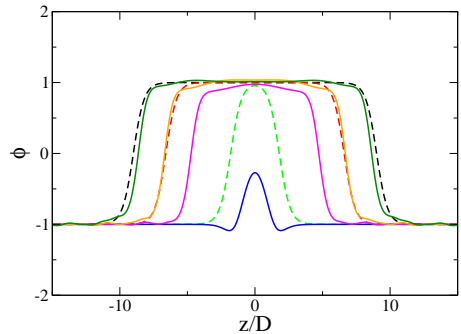
\caption{The incoming kink profiles before, during and after a kink-antikink
  collision. The curves are equally spaced in time, with dashed lines
  before the collision and full lines after.}
}

Before discussing the behaviour of fermions on kinks we first want to describe how kinks interact
with antikinks, and also how they interact with the $\pm BC$ boundaries.
The dynamics of kink-antikink collisions have been studied by a number of authors
\cite{Anninos:1991un,Silveira:1988kc,Belova:1985fg,Campbell:1983xu,Takamizu:2004rq}
revealing the rich structure of behaviour at small impact speeds, $v$. At large speeds, $v\;\gsim\;0.2$,
the kinks simply collide a single time and bounce away to infinity. As we go to smaller speeds then
the kinks may have multiple collisions before travelling off to infinity, or they may simply
annihilate. While extending the analysis of \cite{Gibbons:2006ge} for the fermions involved
in such collisions we shall concentrate on the simplest range of speeds where there is
a single collision, Fig. \ref{fig:confKAK}.

\FIGURE{
\epsfig{file=./pictures/KB_NR_csample.eps,width=7cm,clip}
\epsfig{file=./pictures/KBp_NR_csample.eps,width=7cm,clip}
\label{fig:confs}
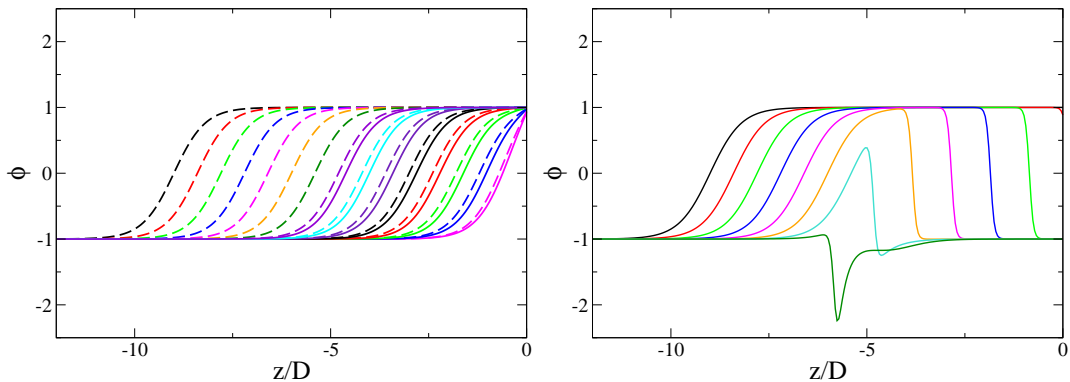
\caption{The incoming kink profiles before and after the collision
  with a $-BC$ boundary (left). The curves are equally spaced
  in time, dashed before the collision, full after. The kink enters the boundary and comes
  back out unscathed. In contrast, the $+BC$ boundary (right) decays straight away and
  emits an antikink which prevents the incoming kink from reaching the
  boundary.}
}

In order to capture the dynamics of brane-boundary collisions, Antunes {\it et al} \cite{Antunes:2003kh}
modelled the system with a scalar field and, in our language, collided kinks against
$ \pm BC$-like boundaries. They
discovered that the kink is temporarily absorbed into the boundary, but re-emerges having lost
a (model-independent) fraction of its kinetic energy of $\sim
63\%$. The boundary conditions used were in fact slightly different
from ours, in that they made the replacement
\be
W'(\phi) = \pm \sqrt{\frac{\lambda}{2}}(\phi^{2}-1)\,\,\rightarrow\,\,|\sqrt{\frac{\lambda}{2}}(\phi^{2}-1)|,
\ee
hence having $-BC$ between the two potential minima
and $+BC$ outside. This fix is responsible for the reported energy loss at collision. 

In our simulations we shall be re-visiting this scenario, but using
strict $\pm BC$ boundaries. Then energy
is conserved at collision, taking into account changes in boundary
energy. For $-BC$, the kink reemerging from the boundary has equal and opposite velocity to the incoming one,
Fig. \ref{fig:confs} (left).
In the case of $+BC$
the vev $\phi(z=0)=1$ at the boundary is unstable to decay via the emission of
an antikink, Fig. \ref{fig:confs} (right). This because the energy difference between $+1$ and $-1$ on
the boundary is exactly the energy of an antikink. In the absence of
an incoming kink (or with a kink very far away), the boundary could
remain in the unstable vacuum, but as the kink approaches, the
exponential tail hits the wall and causes the vacuum to decay. In practice, this
means that before the kink reaches the boundary, an antikink will be
emitted and collide with the incoming kink. Hence we cannot realise a kink/$+BC$ collision.


\section{Particle number and Bogoliubov coefficients\label{sec:partbogo}}


\subsection{Kink-antikink modes\label{sec:KAK1}}

The general philosophy to calculating particle numbers is to identify the correct vacuum and
creation/annihilation operators. For example, when colliding a kink and antikink we consider
the system in the asymptotic past to be in a vacuum state and we can therefore expand the
fermi wave operator as
\bea
\label{eq:inWaveOperator}
\Psi=a^K\psi^K_{in}+a^{KE}\psi^{KE}_{in}+a^{A}\psi^{A}_{in}+a^{AE}\psi^{AE}_{in}+{\rm continuum}.
\eea
Where the $a$ are the particle operators for: the fermi zero mode on the kink ($K$); 
fermi zero mode on the antikink ($A$); first excited fermi mode on
the kink ($KE$);
first excited fermi mode on the antikink ($AE$). And the $\psi_{in}$ are the (normalized) mode
functions found in section \ref{sec:fermbound} corresponding to kinks and antikinks with
the requisite position and velocity.
If there is more than one excited fermion mode then they may also be included in an obvious way.
To an observer in the asymptotic future there will be a similar expansion, only now they
will use a different set of particle operators,
\bea
\label{eq:outWaveOperator}
\Psi=b^K\psi^K_{out}+b^{KE}\psi^{KE}_{out}+b^{A}\psi^{A}_{out}+b^{AE}\psi^{AE}_{out}+{\rm continuum}.
\eea
Again, the mode functions $\psi_{out}$ are the mode functions corresponding to the kink, antikink with
the appropriate position and velocity of the outgoing defects.

From the original action we see that the momentum conjugate to 
the wave operator $\Psi$ is $i\Psi^\dagger$ so by
using the standard equal-time anti-commutation relation
\bea
\{\Psi_\alpha(t,\underline{x}),\Psi^\dagger_\beta(t,\underline{y})\}=\delta_{\alpha,\beta}\delta(\underline{x}-\underline{y}),
\eea
we see that the particle operators obey
\bea
\label{eq:operatorAntiComm}
\{a,a^\dagger\}&=&\{b,b^\dagger\}=1,
\eea
with other anti-commutators vanishing.

Now that we have our wave operator in the asymptotic limits, we need to relate the particle operators
in order to understand how particle numbers are affected. To do this we introduce the Bogoliubov coefficients
in the standard way, which relate the mode functions $\psi_{in}$ to $\psi_{out}$ in the asymptotic future.
In the following expressions the mode functions $\psi_{in}$ are the
time-evolved mode functions from  (\ref{eq:inWaveOperator})
evaluated in the asymptotic future,
\bea
\label{eq:bogoExpansion}
\psi^{K}_{in}&=& \alpha_{K}\psi^{K}_{out} +\beta_{K}\psi^{KE}_{out}
                +\gamma_{K}\psi^{A}_{out}+\delta_{K}\psi^{AE}_{out}
             ~+~{\rm  continuum},\\\nonumber
\psi^{KE}_{in}&=& \alpha_{KE}\psi^{K}_{out}+ \beta_{KE}\psi^{KE}_{out}
                 +\gamma_{KE}\psi^{A}_{out}+\delta_{KE}\psi^{AE}_{out}
             ~+~{\rm  continuum},\\\nonumber
\psi^{A}_{in}&=& \alpha_{A}\psi^{K}_{out}+ \beta_{A}\psi^{KE}_{out}
                +\gamma_{A}\psi^{A}_{out}+\delta_{A}\psi^{AE}_{out}
             ~+~{\rm  continuum},\\\nonumber
\psi^{AE}_{in}&=& \alpha_{AE}\psi^{K}_{out}+ \beta_{AE}\psi^{KE}_{out}
                 +\gamma_{AE}\psi^{A}_{out}+\delta_{AE}\psi^{AE}_{out}
             ~+~{\rm  continuum},
\eea

The $\alpha_{i}$,
$\beta_{i}$, $\gamma_{i}$, $\delta_{i}$ are the Bogoliubov
coefficients, which can be extracted by taking the inner products according to (\ref{eq:fermiNorm})
\bea
\alpha_{K}=(\psi^{K}_{in},\psi^{K}_{out}),&\quad&
\beta_{K}= (\psi^{K}_{in},\psi^{KE}_{out}),\\
\gamma_{K}=(\psi^{K}_{in},\psi^{A}_{out}),&\quad&
\delta_{K}=(\psi^{K}_{in},\psi^{AE}_{out}),
\eea
and similarly for $\psi_{KE}$, $\psi_{A}$, $\psi_{AE}$, $\psi_{B}$. In this way we are able to calculate
all of the Bogoliubov coefficients for a given simulation.

The number operator takes the standard form,
\bea
\hat N&=&\half\int dx(\Psi^\dagger\Psi-\Psi\Psi^\dagger),
\eea
which we may write, using (\ref{eq:inWaveOperator}),
(\ref{eq:outWaveOperator}), (\ref{eq:operatorAntiComm}), in terms of the particle operators
\bea
\hat N&=&\hat n^K+\hat n^{KE}+\hat n^A+\hat n^{AE},\\
\hat n^K&=&{\cal O}^{K\dagger}{\cal O}^K-\half,\;~~~
\hat n^A={\cal O}^{A\dagger}{\cal O}^A-\half,\\
\hat n^{KE}&=&{\cal O}^{KE\dagger}{\cal O}^{KE}-\half,\;~~~
\hat n^{AE}={\cal O}^{AE\dagger}{\cal O}^{AE}-\half,
\eea
where ${\cal O}$ represents either the $a$ or $b$ particle operator depending on whether we
are looking at the past or future respectively.
We define the vacuum relative to the initial state as
\bea
\forall~ i,j,~~~a^{i}|0\rangle=0,
\eea
and the particle states as
\bea
  |K000\rangle=a^{K\dagger}|0\rangle,
~~|0A00\rangle=a^{A\dagger}|0\rangle,
~~|00KE0\rangle=a^{KE\dagger}|0\rangle,
~~|000AE\rangle=a^{AE\dagger}|0\rangle.
\eea
We can see from this that, as described in \cite{Jackiw:1975fn}, the vacuum states have fermion
number $-\half$ whilst the excited states have fermion number $+\half$.

We are now in a position to express the asymptotic-future particle operators, $b$, in terms
of the asymptotic past particle operators, $a$, by comparing (\ref{eq:inWaveOperator}) 
and (\ref{eq:outWaveOperator}) and using (\ref{eq:bogoExpansion}),
\bea
b^{K}&=& \alpha_{K}a^{K}+\alpha_{KE}a^{KE}+\alpha_{A}a^{A}+\alpha_{AE}a^{AE},\\
b^{KE}&=& \beta_{K}a^{K}+ \beta_{KE}a^{KE}+ \beta_{A}a^{A}+ \beta_{AE}a^{AE},\\
b^{A}&=& \gamma_{K}a^{K}+\gamma_{KE}a^{KE}+\gamma_{A}a^{A}+\gamma_{AE}a^{AE},\\
b^{AE}&=&\delta_{K}a^{K}+\delta_{KE}a^{KE}+\delta_{A}a^{A}+\delta_{AE}a^{AE}.
\eea

For simplicity, we will focus our interest on evaluating the
expectation values in the state corresponding to a kink with a zero mode fermion
colliding with a vacuum antikink.
\be
n^{K}=\langle K000|b^{K\dagger}b^K|K000\rangle-1/2,
\ee
and similarly for $n^{KE}$, $n^{A}$, $n^{AE}$, $n^{B}$. These then
represent the fermion occupation numbers in the respective modes after
the collisions, assuming that the initial state has only excitations
(particles) in the kink fermion groundstate. This state is annihilated by all the $a^{KE/A/AE}$, leaving us
to calculate only
\bea
n^{K}+1/2 &=&
\left(n^{K}_{0}+1/2\right)|\alpha^{K}|^{2},\\
n^{KE}+1/2 &=&
\left(n^{K}_{0}+1/2\right)|\beta^{K}|^{2},\\
n^{A}+1/2 &=&
\left(n^{K}_{0}+1/2\right)|\gamma^{K}|^{2},\\
n^{AE}+1/2& =&
\left(n^{K}_{0}+1/2\right)|\delta^{K}|^{2},
\eea
where we have generalised to an initial state with fermions only in
the K mode, but with an arbitrary particle number $n^{K}_{0}$.
We will not be concerned with the normalisation of the state, but
simply compute the Bogoliubov coefficients $\alpha_{K}$,
$\beta_{K}$, $\gamma_{K}$, $\delta_{K}$. Below we will suppress the label $K$.


\subsection{Kink-boundary modes\label{sec:KB1}}

For the collision of a kink on a $-BC$, we now expand the wave operator as
\bea
\label{eq:inWaveOperatorBdry}
\Psi=a^K\psi^K_{in}+a^{KE}\psi^{KE}_{in}+a^{B}\psi^{B}_{in}+{\rm continuum}.
\eea
and
\bea
\label{eq:inWaveOperatorBdryout}
\Psi=b^K\psi^K_{out}+b^{KE}\psi^{KE}_{out}+b^{B}\psi^{B}_{out}+{\rm continuum},
\eea
where the index $B$ refers to the boundary zero mode.
The expansion of the mode functions proceeds as before, with the addition of a boundary mode function,
\bea
\label{eq:bogoExpansionBdry}
\psi^{K}_{in}&=& \alpha_{K}\psi^{K}_{out} +\beta_{K}\psi^{KE}_{out}                
                +\xi_K\psi^B_{out}
             ~+~{\rm  continuum},\\
\psi^{B}_{in}&=& \alpha_{B}\psi^{K}_{out}+ \beta_{B}\psi^{KE}_{out}
+\xi_{B}\psi^B_{out}
             ~+~{\rm  continuum},
\eea
and if we send in a kink with non-vanishing occupation number $n_K^0$ for the
zero-mode fermion, then we find that the boundary number operator
in the asymptotic future is
\bea
n^{B}+1/2& =&
\left(n^{K}_{0}+1/2\right)|\xi^{K}|^{2}.
\eea


\section{Kink-antikink collisions\label{sec:KAK}}

\FIGURE{
\epsfig{file=./pictures/KK_NR_conf.eps,width=7cm,clip}
\epsfig{file=./pictures/KK_NR_bsample.eps,width=7cm,clip}
\caption{Left: The scalar field (black) and the fermion $K$ mode components
  $\psi_{1}$ (red) and $\psi_{2}$ (blue) before (dashed) and after (full) the
  collision. The initial velocity was $v=0.6$, and the coupling
  $g=2$. Right:  The overlap of the incoming K mode after the collision with
  the outgoing kink K and antikink A modes ($\alpha$ and
  $\gamma$), and the KE and AK modes ($\beta$ and $\delta$).}
\label{fig:KAKconf}
}
The first set of results that we shall present is an extension of the
study in \cite{Gibbons:2006ge}, where the Bogoliubov coefficients of the
fermion zero mode were calculated for  kink/antikink collisions. Here
we also present data for the excited fermion bound state, as well as
observing the dependence on collision speed.

The collisions were performed by initially placing a kink
and a antikink a distance $30 v$ apart. We then boosted them with
velocity $v$ and $-v$ respectively, as described
above. Fig. \ref{fig:KAKconf} (left)
shows the profile of the scalar field and
fermion modes before and after the collision, for the case of $v=0.6$, $g=2$.
Both before and after the collision, the fermion modes are well
localised around the kink and antikink. 

We then calculate the Bogoliubov coefficients in time,
Fig. \ref{fig:KAKconf} (right). Before the collision at $Dt\simeq 15$
the fermion mode is the kink K mode, and so $|\alpha|^{2}=1$. during
the collision all bets are off; in particular we are not able to
assign velocities to individual kinks. Already at time $Dt=20$, the
kink-antikink pair have disentangled themselves, and a final
Bogoliubov coefficient had been established. Although there is some
residual oscillation even at late times, we assign final values at
$Dt=50$.

It is worth noting, that because the phase $\varphi$ in Eqs. (\ref{eq:Emodes1},
\ref{eq:Emodes2}) is undetermined, we instead keep the whole
time-dependence $\omega t+\varphi$ fixed when computing the
overlap. This means that the computed $\beta$ and $\delta$ are oscillating
functions in time, Fig. \ref{fig:KAKconf} (right), and we should use the amplitude of this
oscillation as the Bogoliubov coefficient.
\FIGURE{
\epsfig{file=./pictures/KK_NR_alpha.eps,width=7cm,clip}
\epsfig{file=./pictures/KK_NR_gamma.eps,width=7cm,clip}
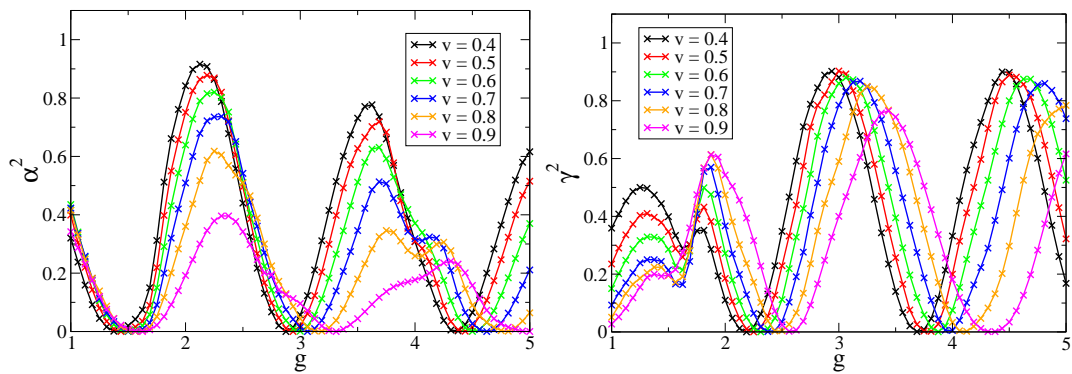
\caption{Left: The velocity and coupling dependence of the Bogoliubov
  coefficients of the incoming $K$ mode with the outgoing $K$
  mode. Right: The same for the overlap on the $A$
  mode. Errorbars reflecting the residual oscillation (see text) are roughly the
  size of the symbols.}
\label{fig:bogo2}
}
Taking this into account, the Bogoliubov coefficients can be read off
with an accuracy of in most cases better than
$0.01$. Fig. \ref{fig:bogo2} shows the $g$ and $v$ dependence of
$\alpha^{2}$ and $\gamma^{2}$, the overlap with the outgoing K and A modes.

As noted in \cite{Gibbons:2006ge}, the coupling dependence approximately
follows a $a+b\sin(cg+d)$ form, although the amplitude decreases
somewhat with $g$, especially for $\alpha$. We will not attempt to fit
this behaviour numerically, but simply note some qualitative points of
interest.
$\alpha^{2}$ and $\gamma^{2}$ are anti-correlated with the same period
in $g$, and the bulk of the fermion
number ends up in these lowest energy modes after the
collision. However, as $v$ is increased, fermion number is lost from
these modes, in particular from $\alpha^{2}$. For small $g$, the 
Bogoliubov coefficients take a very long time to settle and seem 
to decrease continuously until they do so. 
\FIGURE{
\epsfig{file=./pictures/KK_NR_beta.eps,width=7cm,clip}
\epsfig{file=./pictures/KK_NR_delta.eps,width=7cm,clip}
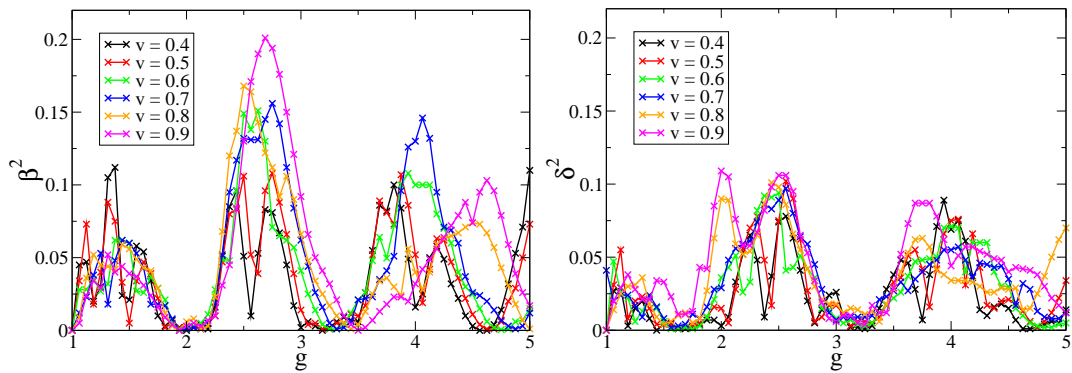
\caption{Left: The velocity and coupling dependence of the Bogoliubov
  coefficients of the K mode with the outgoing kink KE
  mode. Right: The same for the overlap on the antikink AE mode.}
\label{fig:bogo3}
}
Fig. \ref{fig:bogo3} shows the corresponding overlap with the KE and
AE modes, $\beta^{2}$ and $\delta^{2}$. These are strongly correlated
with each other and seem to have the same period in $g$ as K/A mode
coefficients, but with a phase shift of $\pi/2$. The fermion number
taken away in these modes is much smaller, but it is interesting that
up to $10$ to $20$ percent can end up here. There is a mild increase with
$v$ consistent with the decrease in $\alpha^{2}$ and $\gamma^{2}$.
\FIGURE{
\epsfig{file=./pictures/KK_NR_sum.eps,width=7cm,clip}
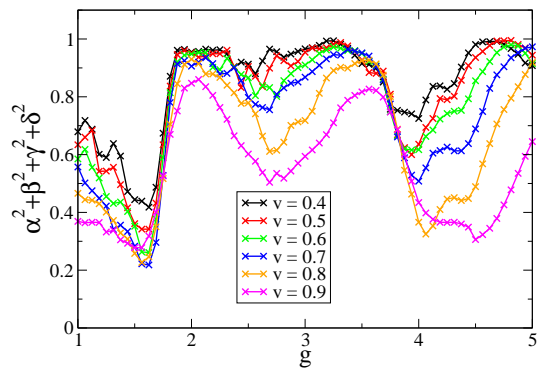
\caption{The sum of Bogoliubov overlaps on all the bound states 
$|\alpha|^{2}+|\beta|^{2}+|\gamma|^{2}+|\delta|^{2}$.
The deviation from 1 is the amount of fermion number carried away as radiation.}
\label{fig:bogo4}
}
Any missing fermion number must then be transfered to modes which we
do not take into account, radiation or additional time-dependent bound
states. We quantify this by calculating the sum of the coefficients
$|\alpha|^{2}+|\beta|^{2}+|\gamma|^{2}+|\delta|^{2}$, shown in 
Fig. \ref{fig:bogo4}. The loss to radiation has some non-trivial
dependence on $g$, and increases with $v$, presumable because there is
then energy available to excite higher energy modes. It is striking
that up to $60$ percent of the fermion number can be lost to radiation
in this way.


\section{Kink-boundary collision\label{sec:KB}}

The second set of results in this paper are those associated with
a brane colliding into a boundary, rather than another brane.
as such, we 
now collide a single kink onto $-BC$ boundary and see
to what extent a fermion originally localised on the kink will
stick to the boundary.
Fig. \ref{fig:KBall} (left) shows the field profiles initially and at late
times after the collision. Again, the fermions remain nicely localised around
the kink, but after the collision also at the boundary.

We shall again consider only the case where the kink starts with a fermion
zero mode,
and calculate the Bogoliubov coefficient for
finding a fermion in the outgoing K ($|\alpha|^{2}$),
the outgoing KE ($|\beta|^{2}$) and the boundary
B ($|\xi|^{2}$) modes. Fig. \ref{fig:KBall} (right) show the
coefficients in time, again with $\alpha^{2}=1$ until the collision at
$Dt\simeq 15$. After another transient collision stage, the
coefficients take on definite values, and we end the simulation at
$Dt=50$. The results concerning fermions radiating into the bulk are somewhat different
to the kink/antikink collisions. As shown in Fig. \ref{fig:KBbogoSum} we see that the
sum representing bound state fermions, $|\alpha|^{2}+|\beta|^{2}+|\xi|^{2}$,  is very close to unity.
This implies that very little of the fermions end up in the bulk, with larger collision speeds
producing more bulk fermions as one would expect.
\FIGURE{
\epsfig{file=./pictures/KB_NR_conf.eps,width=7cm,clip}
\epsfig{file=./pictures/KB_NR_bsample.eps,width=7cm,clip}
\caption{Left: The scalar field (black) and the fermion $K$ mode components
  $\psi_{1}$ (red) and $\psi_{2}$ (blue) before (dashed) and after (full) the
  collision. The initial velocity was $v=0.6$, and the coupling
  $g=2$. Note the non-zero boundary mode contribution to the far right. Right: The overlap of the K mode after the collision with
  the outgoing kink K and KE modes ($\alpha$ and $\beta$) and the boundary mode B
  ($\xi$).}
\label{fig:KBall}
}
We again determine the dependence on $g$ and $v$, shown in
Fig. \ref{fig:KBbogo1}. In this case $\alpha^{2}$ is almost exactly anti-correlated
with $\xi^{2}$. Most fermion number is transfered to the boundary for
small and large values of $g$, with the maxima moving down and up ,
respectively as $v$ is increased. For $v=0.9$, the kink retains its
fermion, at least in the range of couplings used here.
\FIGURE{
\epsfig{file=./pictures/KB_NR_alpha.eps,width=7cm,clip}
\epsfig{file=./pictures/KB_NR_xi.eps,width=7cm,clip}
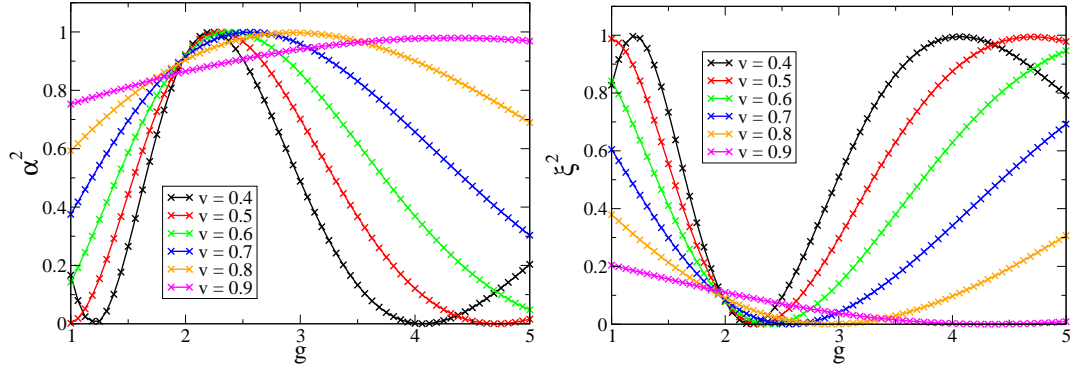
\caption{The Bogoliubov coefficients of the incoming K mode unto the
outgoing K mode, $|\alpha|^{2}$, (left) and B mode, $|\xi|^{2}$, (right).}
\label{fig:KBbogo1}
}
\FIGURE{
\epsfig{file=./pictures/KB_NR_sum.eps,width=7cm,clip}
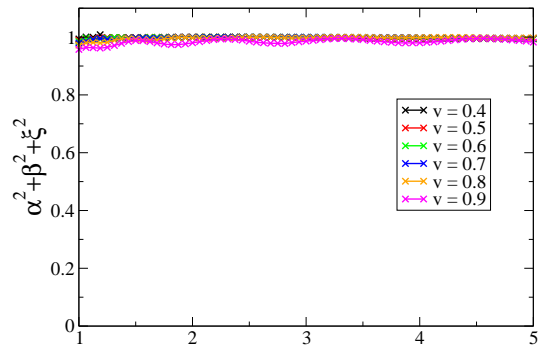
\caption{The sum, $|\alpha|^{2}+|\beta|^{2}+|\xi|^{2}$, representing the fraction of
fermions that end up in bound states.}
\label{fig:KBbogoSum}
}

In order to understand the kink/antikink results we repeat the approximation developed in
\cite{Gibbons:2006ge}. The approximation is a way of solving (\ref{eq:phieom}), (\ref{eq:eom2})
which writes the fermi field during the collisions as
\bea
\psi_{(1,2)}\simeq A_{(1,2)}f(z),
\eea
where $f(z)$ is some even, normalized function, and the total amplitude is normalized by
$A_{(1)}^2+A^{2}_{(2)}=1$. We then find that (\ref{eq:phieom}), (\ref{eq:eom2}) may be integrated
along the $z$ axis to give
\bea
\label{eq:analBog}
A^2_{(1,2)}\simeq \half\left(1+\sin(2g\phi_c\Delta t)\right).
\eea
In this expression we have a representative value for $\phi$ during the collision, $\phi_c$,
and a collision timescale $\Delta t$. In this way Gibbons {\it et al} were able to explain
the sinusoidal behaviour of the Bogoliubov coefficients when a kink collides with an antikink.
Unfortunately, the case of a kink colliding with a boundary does not succumb to the same
analysis, largely because the fermion mode functions have rather different $z$-dependence owing
to one of them being forced to vanish at the boundary. However, from the form of the Bogoliubov
coefficients it is tempting to speculate that a relation similar to (\ref{eq:analBog}) holds.
A possible explanation for the longer ``wavelengths'' in Fig. \ref{fig:KBbogo1} could then be that
during the kink/boundary collision the value of $\phi$ changes very little from its vacuum value,
while in the kink/antikink collisions one finds that $\phi$ overshoots the
vacuum by some amount depending on collision speed.


\section{Conclusion\label{sec:conclusion}}

In summary, we have performed a detailed numerical study of fermion transfer in
kink-antikink and kink-boundary collisions.

In kink-antikink collisions, we confirm the findings 
of Gibbons {\it et al} \cite{Gibbons:2006ge} that although the scalar field
kinks bounce off each other in an elastic way, fermions initially in
the K mode on the kink will be distributed on the K, A, and radiation modes.
As an extension of their work we also included the first fermion
excited modes, KE, AE,
finding that these modes also gets excited, and presented data for a wide range
of collision speeds.
The distribution is very sensitive to the value of
the coupling $g$, and hence the mass of the fermions $g\phi$. For small
incident velocities $v$, most of the fermion number ends up in the K
and A modes, and these are anti-correlated in $g$. As the velocity is
increased, more and more fermion number is transfered from the K/A
modes to the KE and AE, but also delocalised radiation modes, and up
to $60$ percent can be ``lost'' to the bulk.

When colliding kinks on a -BC boundary, we found again that the kink
is reflected elastically, but that a significant amount of fermion
number can stick on the boundary. In this case, the K and B modes are
anticorrelated in $g$, but in contrast to the kink-antikink collisions, very
little fermion number is transfered to the KE mode or radiation. For
the values of $g$ employed here, low $v$ favours transfer to the
boundary, whereas for high $v$ the fermions stay on the kink and are
carried away from the boundary again.

We in fact also solved for the evolution of the initial B, A, KE, and AE
  modes, and unsurprisingly found that fermions initially in these
  modes are also distributed on all modes after collision. In
  particular, a fermion localised on the boundary can be carried away
  by a kink bouncing on this boundary. For brevity and to focus on our
  main aim, we did not carry out a detailed exposition of all mode
  combinations. It is however straightforward to do so.

We found that the numerical implementation of the $\pm BC$
  boundary conditions require some care, and that discretisation
  errors can be significant. We dealt with this by using iterative
  methods, higher order derivatives and rather fine lattices, $D dr=0.0025$.

One could consider including more and more fermion modes in the spectrum,
  in order to track the ``lost'' fermions. In the end, this would lead
  to a full quantum (Hartree) treatment \cite{Aarts:1998td}, and one
  would be able to include the back-reaction on the scalar
  self-consistently. For the purpose of this paper, we found the K, A, KE, AK
  and B modes to be sufficient.

We started by arguing that our 1+1 model could be lifted to give results for
  a 4+1 spacetime. While this is true, one has ignored any effects that
  depend on the directions within the brane. In particular we have not presented
  any results about the $\underline k$ dependence of the Bogoliubov coefficients
  (where $\underline k$ refers to the wave-number in the brane world-volume.)
  We hope to present results on this in a future publication.

Another natural extension of this work is to consider more than
  one species of fermions and/or more scalars to include effects like C
  and CP violation in the scalar-fermion interaction. This would for
  instance be relevant to electroweak baryogenesis, where at a first
  order phase transition, a domain
  wall sweeps through a plasma, reflecting fermions off it in a
  CP-violating way.


\subsection*{Acknowledgments}
We thank Kei-Ichi Maeda for stimulating discussions. P.M.S is supported by PPARC
and A.T. is supported by PPARC Special Programme Grant {\it``Classical
  Lattice Field Theory''}. We gratefully acknowledge the use of the UK
National Cosmology Supercomputer, Cosmos, funded by PPARC, HEFCE and Silicon
Graphics.

\bibliographystyle{JHEP}

\end{document}